\documentclass[]{aa}
\usepackage{graphics}
\begin{document}

\def\bib{\bibitem{}}

\def\vd{{\vec d}} \def\ve{{\vec e}} \def\vu{{\vec u}} \def\vl{{\vec l}} \def\vx{{\vec x}} \def\vk{{\vec
k}} \def\vgam{{\vec{\gamma}}} \def\vxi{{\vec{\xi}}} \def\mD{{\cal{D}}} \def\d{{\rm d}}
\def\i{{\rm i}} \def\veps{{\vec \epsilon}} \def\eps{\epsilon}
\def\gam{\gamma}

\def\bold{}

%\thesaurus{Sect.02 (12.03.4; 12.07.1; 12.12.1)}

\title{Patterns in the  weak shear 3-point correlation function}
\author{F. Bernardeau\inst{1} \and L. van Waerbeke\inst{2,3} \and
Y. Mellier\inst{2,4}} \institute{Service de Physique Th{\'e}orique,
CEA/DSM/SPhT, Unit{\'e} de recherche associ{\'e}e au CNRS, CEA/Saclay
91191 Gif-sur-Yvette c{\'e}dex, France \and Institut d'Astrophysique
de Paris, 98 bis bd Arago, 75014 Paris, France \and Canadian
Institute for Theoretical Astrophysics, 60 St George Str., M5S
3H8, Toronto, Canada \and LERMA, Observatoire de Paris, 61 avenue
de l'Observatoire, 75014 Paris, France }

\date{Received / Accepted }

\abstract{We explore the scale and angular dependence of the
cosmic shear three-point correlation function. The shear field is
found to have a much more complex three-point function than the
convergence field, but it also exhibits specific motifs which draw
signatures of gravitational lensing. Exact shear patterns are
inferred analytically for some geometrical shear triplets
configurations, when simple interpretations can be derived. {A
more complete description of their geometrical properties has then
been carried out} from ray-tracing numerical simulations and
simple models. These patterns reveal  typical  features that can
be utilized for non-gaussian signal detection in cosmic shear
surveys. We test the robustness of these properties against
complex noise statistics and non-trivial survey topologies. From
these conclusive checks, we predict that the VIRMOS-DESCART survey
should allow a detection of a non-gaussian signal with a
comfortable significance for a low matter density Universe.

}

\maketitle

\section{Introduction}

The detections of cosmic shear signal (van Waerbeke et al.
2000, Bacon et al.  2000,  Wittman et al.   2000, Kaiser et  al.
2000, Maoli et al 2001) opened the analysis of  large-scale mass
 distribution in the
Universe to weak lensing  surveys. As the surveys sizes progressively
increase, the noise and systematics are reduced to  reasonable levels,
which permits to constrain cosmological parameters from the amplitude
and the shape of the shear two-point correlation function (Maoli et
al. 2001, van Waerbeke et al. 2001b). However, although it is weakly
sensitive to the cosmological constant $\Omega_{\Lambda}$ (Bernardeau, van
Waerbeke \& Mellier 1997),   the two-point function is a degenerate
combination of the matter density $\Omega_{{\rm m}}$ of the Universe and the
amplitude of the power spectrum $\sigma_8$ (Villumsen 1996, Jain \&
Seljak 1997).  \\ This degeneracy can be broken from the angular
dependence of the cosmic shear amplitude (Jain \& Seljak 1997), but it
relies on the prior knowledge of the shape of the mass power spectrum.
An alternative is to directly probe weak shear maps,  which contain
non-gaussian features that can be used to derive $\Omega_{{\rm m}}$ with the
only assumption that initial conditions were gaussian (Bernardeau, van
Waerbeke, Mellier 1997, van Waerbeke, Bernardeau, Mellier 1999) - this
later assumption being eventually testable in the data set itself .
\\ So far, all the theoretical predictions regarding non-gaussian
features are based on a critical reconstruction process of either a
convergence map (filtered for instance with a top-hat window function)
or an aperture mass map (Kaiser et al. 1994, Schneider 1996, Schneider
et al. 1998, Bernardeau \& Valageas 2000).  Unfortunately, the
panoramic reconstruction of mass maps from real data turned out to be
considerably more difficult than expected.  The masking process
discussed by van Waerbeke et al. 2000 produces patchy surveys with a
non-trivial topology and inhomogeneous noise. The resulting mass maps
have poorly understood  statistical properties which is practically
difficult to handle with confidence. We therefore explore another
option which uses direct signatures of gaussian effects in shear map
patterns.

Shear pattern study is a priory more difficult because the third order
moment of the local shear vanishes for obvious symmetry reasons. It is
therefore necessary to seek for peculiar configurations (or
geometries) of shear triplets for which a significant signal is
expected. This is the aim of this paper. In the next section, we
present the exact analytical results obtained for specific geometries,
their interpretation and the results obtained from ray-tracing
simulations which exhibit the complete three-point function
patterns. The detection of these features in these simulations is
presented in section 3. These results suggest a detection strategy in
cosmic shear surveys which is put forward and tested with mock
catalogs in section 4. We checked carefully that the masks do not
compromise the possibility  of detecting the effect. Finally, we
estimate the expected error level of such measurements in the
available data sets.

\section{Theoretical insights}

\subsection{The local shear statistical properties}

In the following we assume  the shear signal is entirely due to
cosmic shear effects, and there is no so-called $B$-mode produced
by lens-lens coupling,  clustering or any other contamination. Later
in the paper (in Section 4) we study the impact of residual
systematics. This hypothesis naturally leads to derive the shear field
$\vgam$ from a potential field which can be written, in Fourier space,
\begin{equation}
\vgam(\vx)=\int\d^2\vl\,\kappa_{\vl}\,\exp(\i\vl.\vx)\,\vu(\theta_{\vl})
\end{equation}
with
\begin{equation}
\vu(\theta_{\vl})=\left(
\begin{array}{c}
\cos(2 \theta_{\vl})\\ \sin(2 \theta_{\vl})
\end{array}
\right),
\end{equation}
where $\kappa_{\vl}$ is the Fourier transform of the local
convergence map, {the components of $\vl$ being
$l\cos(\theta_{\vl})$ and $l\sin(\theta_{\vl})$.} We also assume
the validity of the small angle approximation, making the
spherical harmonic decomposition unnecessary (Stebbins 1996).

The statistical properties of the variable $\kappa _{\vl}$ can be
inferred from those of the projected density contrast. These
derivations have been studied in details in the many papers mentioned
in the introduction. Using    the small angle approximation (Limber
1954, applied here in Fourier space, Kaiser 1992, 1998, Bernardeau
1995), they give the expression of the convergence power spectrum
$P_{\kappa}(l)$, defined as,
\begin{equation}
\langle \kappa_{\vl}\kappa_{\vl'}\rangle= \delta_{\rm
Dirac}(\vl+\vl')\,P_{\kappa}(l)
\end{equation}
as an integral over the line-of-sight of the 3D matter density
contrast power spectrum $P_{\delta}(k)$,
\begin{equation}
P_{\kappa}(l)=\int_0^{\chi_{{\rm CMB}}} {\d\chi\over
\mD^2}\,w^2(\chi)\,P_{\delta}\left(k\over\mD\right)
\end{equation}
where $w(\chi)$ is the lensing efficiency function ($\chi$ is the
radial distance, $\mD$ is the comoving angular distance).  It depends on the
cosmological parameters and on the source redshift distribution
function,  $n(z_s)$:
\begin{equation}
w(\chi)={3\over 2}\Omega_{{\rm m}}\int\d
z_s\,n(z_s){\mD(\chi_s-\chi)\,\mD(\chi)\over a\,\mD(\chi_s)}
\end{equation}
where $\chi_s$ is the radial distance to the source plane and when
the distances are expressed in units of $c/H_0$. The
phenomenological properties of the convergence field, or similarly
the shear field, will be determined by both the shape of the
efficiency function and the properties of the cosmic density
field.

Likewise, the convergence three-point function can be derived from the
cosmic matter three-point function. The small angle approximation can
indeed be used again at this level (Bernardeau 1995, Bernardeau et
al. 1997), leading to the expression of the $\vl$-space three
point function
\begin{equation}
\langle \kappa_{\vl_1}\kappa_{\vl_2}\kappa_{\vl_3}\rangle= \delta_{\rm
Dirac}(\vl_1+\vl_2+\vl_3)\,B_{\kappa}(\vl_1,\vl_2,\vl_3)
\end{equation}
with
\begin{equation}
B_{\kappa}(\vl_1,\vl_2,\vl_3)= \int_0^{\chi_{{\rm CMB}}}{\d\chi\over
\mD^2}\,w^3(\chi)\,B_{\delta}\left({\vl_1\over\mD},{\vl_2\over\mD},{\vl_3\over\mD}\right)
\end{equation}
where $B_{\delta}(\vk_1,\vk_2,\vk_3)$ is the 3D matter density
contrast bispectrum:
\begin{equation}
\langle\delta_{\vk_1}\delta_{\vk_2}\delta_{\vk_3}\rangle= \delta_{\rm
Dirac}(\vk1+\vk_2+\vk_3)\,B_{\delta}(\vk_1,\vk_2,\vk_3).
\end{equation}

The bispectrum $B_{\delta}$ is usually parametrized as follows
\begin{equation}
B_{\delta}(\vk_1,\vk_2,\vk_3)=Q(\vk_1,\vk_2)P_{\delta}(k_1)P_{\delta}(k_2)
+{\rm sym.} \ ,
\end{equation}
where $Q$ is an homogeneous function of the wave vectors $\vk_1$
and $\vk_2$. For instance, in the quasi-linear regime it is given
by,
\begin{equation}
Q(\vk_1,\vk_2)={10\over 7}+{\vk_1.\vk_2\over k_1^2}+ {\vk_1.\vk_2\over
k_2^2}+{4\over 7}{(\vk_1.\vk_2)^2\over k_1^2\,k_2^2}\ ,
\end{equation}
whereas in the nonlinear regime $Q$ is often assumed to be a pure
number (that depends however on the cosmological models). Several
phenomenological models have been proposed to describe the behavior of
the three-point matter correlation function, from the quasi-linear
regime to the nonlinear regime (the so-called EPT, Colombi et al. 1997
or its extensions with the HEPT, Scoccimarro \& Frieman 1999). In
particular Scoccimarro \& Couchman (2001) have
   shown that the matter three-point function
could be described by,
\begin{eqnarray}
Q(\vk_1,\vk_2)&=&a(k_1,k_2)+b(k_1,k_2)\left({\vk_1.\vk_2\over
k_1^2}+{\vk_1.\vk_2\over k_2^2}\right)+\nonumber\\
&&+c(k_1,k_2){(\vk_1.\vk_2)^2\over k_1^2\,k_2^2},\label{heptdelta}
\end{eqnarray}
where $a$, $b$ and $c$ depend on scale in a known way.

The properties of the 3D matter three-point function actually extend
to the three-point correlation function of the convergence field;
e.g. one expects to have
\begin{equation}
B_{\kappa}(\vl_1,\vl_2,\vl_3)=
Q_{\kappa}(\vl_1,\vl_2)P_{\kappa}(l_1)P_{\kappa}(l_2)+{\rm sym.}.
\label{Qkappa}
\end{equation}
To a large extent the functional dependence of the three-point
function coefficient $Q_{\kappa}$ is left unchanged (T{\'o}th, Holl{\'o}si
\& Szalay 1989, Bernardeau 1995) although its amplitude is
affected by projection effects. This has been investigated in
detail at the level of the convergence reduced skewness, which {is
proportional} to
 the angular
averages of $Q_{\kappa}(\vk_1,\vk_2)$\footnote{In the non-linear
regime one simply has $s_3\approx 3\,Q_{\kappa}$ to a very good
approximation}. Interestingly the reduced skewness, defined as
\begin{equation}
s_3={\langle \kappa^3\rangle\over \langle \kappa^2\rangle^2},
\end{equation}
  depends on the efficiency function amplitude, and therefore
on the cosmological parameter, $\Omega_{{\rm m}}$, but is less sensitive to the
cosmological constant value or the vacuum equation of state (Benabed
\& Bernardeau 2001). For sources at redshift about unity, calculations
give (Bernardeau et al. 1997, Hui 1999, van Waerbeke et al.  2001a),
\begin{equation}
s_3\sim {\Omega_{{\rm m}}^{-0.8}},
\end{equation}
irrespective of  the amplitude of the matter fluctuation
\footnote{The reduced skewness can however depend on the
normalization of the  matter density fluctuations in the
intermediate regime between the quasi-linear and the nonlinear
regime}.  This result actually extents to the amplitude of the
convergence three-point function so that one expects to have,
\begin{equation}
Q_{\kappa}\sim {\Omega_{{\rm m}}^{-0.8}},
\end{equation}
also. In the following the inter-relation between $Q$ and $s_3$ will
be investigated in more details. But it is already obvious that a
measurement of $Q$ would be useful to constrain the
 matter density parameter
of the Universe.

\subsection{The shear three-point function}

The general expression for the shear three-point function can be
inferred from  the $\vl$-space three point function of the
convergence field. {However, because $\vgam$ is a 2 component
pseudo-vector field there are many possible ways to combine shear
triplets. The properties of shear patterns produced by these
triplets
  may be complex.  In order to interpret them in a cosmological
  context it is preferrable to focus first on few of them} that
can be easily investigated with analytical calculations. This is
the case for $\left\langle
\left(\vgam(\vx_1).\vgam(\vx_2)\right)\vgam(\vx') \right\rangle$
for a fixed $\vx_2-\vx_1$ separation which is a vectorial
quantity. This function displays a specific pattern which can be
viewed as a 2D vector field.

\subsubsection{Computation of $\langle \gamma^2(\vx)\vgam(\vx')\rangle$}

\begin{figure*}
\begin{tabular}{ccc}
%\vspace{.2cm}
{\centering\resizebox*{5cm}{!}{\includegraphics{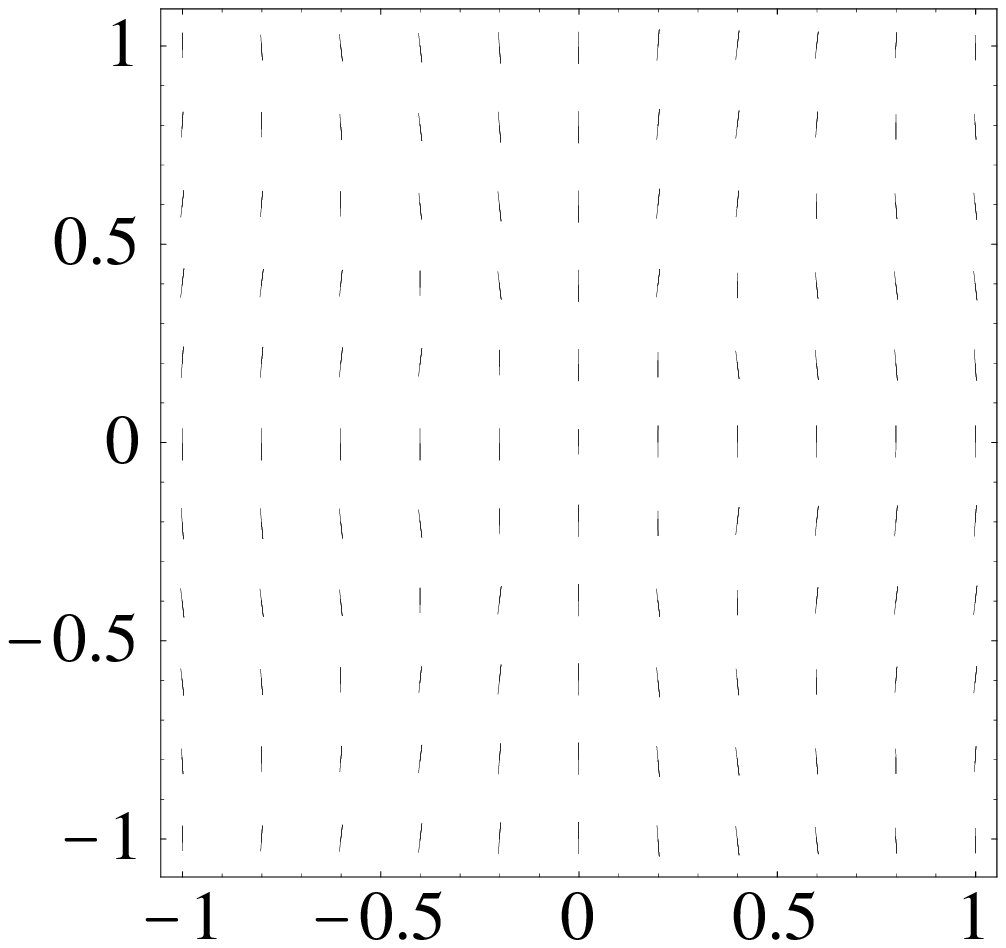}}}
{\centering\resizebox*{5cm}{!}{\includegraphics{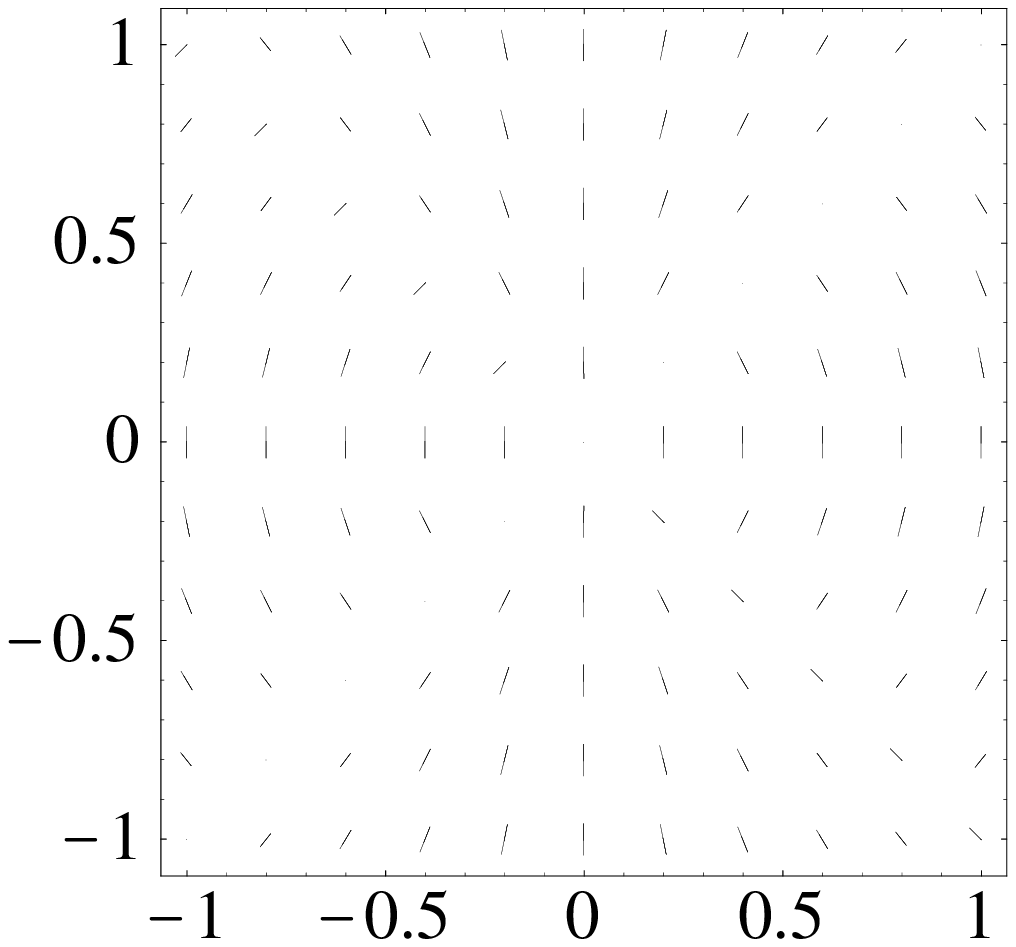}}}
{\centering\resizebox*{5cm}{!}{\includegraphics{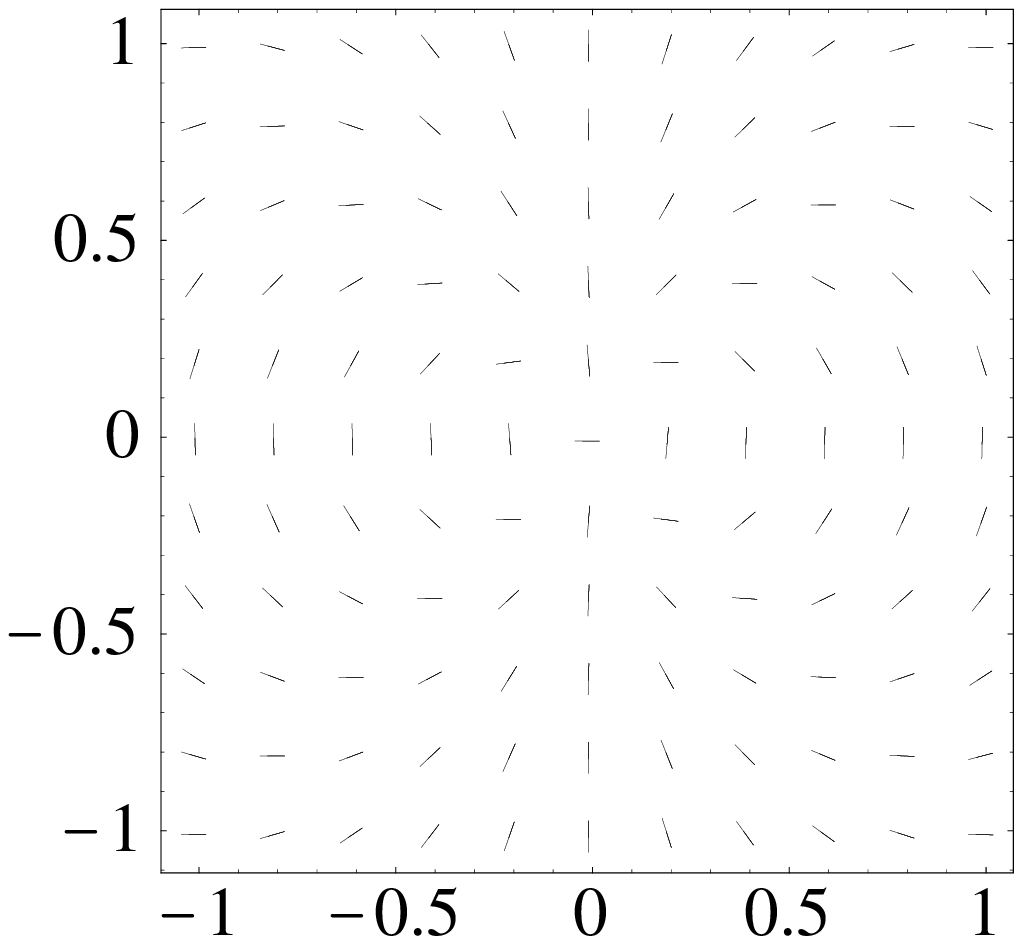}}}
\end{tabular}
\caption{{Shear patterns in the vicinity of one of the pair point
\bold{either $\vx_1$ or $\vx_2$ (positioned at the center of the
plot)} for $n=-1.5$, $n=-1.$ and $n=-0.5$ (from left to right) as
a function of $\vx'$. It illustrates the results of Eq.
(\ref{s3nearpair}). Coordinates are arbitrary.}}
\label{shears3nm1p5}
\end{figure*}

Let start with a simple configuration, when $\vx_1$ and $\vx_2$
are pointing in the same location $\vx$, {for which case we have
$\vgam(\vx_1).\vgam(\vx_2)=\gamma^2(\vx)$}. In this case the
results make sense only when the filtering effects are taken into
account. \bold{Otherwise the result is dominated by the shear
behavior at arbitrarily small scale and depends on the detailed
behavior of the power spectrum and bi-spectrum in the limit $k\to
\infty$; in a regime that might be totally irrelevant for the
observations.} {The shear field is averaged within a smoothing
window, but in order to keep a simple notation, we still denote
$\vgam(\vx)$ the smoothed shear vector at the window location
$\vx$}. For a top-hat window function of radius $\theta_0$, the
three-points function can be written as,
\begin{eqnarray}
\langle \gamma^2(\vx)\vgam(\vx')\rangle&=&\nonumber\\
&&\hspace{-2cm}\int\d^2\vl\,\d^2\vl'\,
W(l\,\theta_0)\,W(\vert\vl+\vl'\vert\theta_0)\,W(l'\,\theta_0)
\,P_{\kappa}(l)\,P_{\kappa}(l')\times\nonumber\\
&&\hspace{-2cm}\,Q_{\kappa}(\vl,\vl')\left[
2\cos(2\theta_{\vl+\vl'}-2\theta_{\vl'})\,e^{\i\vl'.(\vx'-\vx)}\,\vu(\theta_{\vl'})+
\right.\nonumber\\ &&\hspace{0cm}\left.
\cos(2\theta_{\vl}-2\theta_{\vl'})\,e^{\i(\vl+\vl').(\vx'-\vx)}\,\vu(\theta_{\vl'+\vl})
\right] \label{gam2gamp}
\end{eqnarray}
{where $W$ is the $\vl$-space top-hat window function,

\begin{equation}
W(x)={J_1(x)\over x},
\end{equation}
where $J_1(x)$ is the first Bessel function of the first kind.}

{The computation of expression (\ref{gam2gamp}) is still
complicated in general, but it simplifies in the large separation
limit, $\theta_0\ll\vert\vx-\vx'\vert$. In this case it is
possible to express the result in terms of the square root of the
variance of the filtered convergence $\sigma_{\kappa}$ and the
convergence-shear correlation function $\xi_{\kappa\,\gamma}$. }

{In the large separation limit, in the first term of Eq.
(\ref{gam2gamp}), the values of $l'$ that will most contribute are
of the order of $1/\vert\vx-\vx'\vert$ whereas that of $l$ will be
of the order of $1/\theta_0$; in the second term the integration
in $l$ and $l'$ factorize out (in this configuration
$W(\vert\vl'+\vl\vert)$ does not play any role) and both $l$ and
$l'$ are of the order of $1/\theta_0$. As a consequence the first
term of (\ref{gam2gamp}) is of the order of
$\sigma^2_{\kappa}\,\xi_{\kappa\,\gamma}$ whereas the second is of
the order of $\xi^2_{\kappa\,\gamma}$ and is thus
smaller\footnote{{this is a very general scheme for such
computations. It is described in more details in Bernardeau et al.
2001}}. Furthermore, as in the first term $l'\ll l$, it is
possible to expand all quantities in $l'/l$,}
\begin{eqnarray} \cos(2\theta_{\vl+\vl'}-2\theta_{\vl'})&\approx&
1-{l'^2\over2\,l^2}\sin^2(\theta_{\vl}-\theta_{\vl'})+\dots\label{cos2tmtp}\\
W(\vert\vl+\vl'\vert\theta_0)&\approx&
W(l\theta_0)+l\theta_0\,W'(l\theta_0){\vl.\vl'\over l^2}+\dots
\end{eqnarray}
Once a  shape for $Q_{\kappa}$ is given, computing the integral of
the angle  of $\vl$ is straightforward. In general, following the
prescription of Scoccimarro \& Couchman, Eq.  (\ref{heptdelta}),
one can write\footnote{The coefficients $a$, $b$ and $c$ that
appear in (\ref{heptdelta}) and (\ref{heptkappa}) are not
necessarily identical because of the projection effects.}
\begin{equation}
Q_{\kappa}(\vl,\vl')=a(l,l')+b(l,l')\left({\vl.\vl'\over
l^2}+{\vl.\vl'\over l'^2}\right)+c(l,l'){(\vl.\vl')^2\over
l'^2\,l^2} \label{heptkappa}
\end{equation}
from which one gets,
\begin{eqnarray}
\langle \gamma^2(\vx)\vgam(\vx')\rangle&=& 2\pi\,\int l\d l\,\d^2
\vl'\, P_{\kappa}(l)\,P_{\kappa}(l')\,e^{\i\vl'.(\vx'-\vx)}
\times\nonumber\\
&&\hspace{-2cm}\left\{W^2(l\,\theta_0)\,W(l'\,\theta_0)
\left[a(l,l')+{c(l,l')\over2}\right]\right.+\nonumber\\
&&\hspace{-2cm}\left.
{l\theta_0\over2}W(l\theta_0)W'(l\theta_0)W(l'\,\theta_0)b(l,l')\right\}
\,\vu(\theta_{\vl'}). \label{3ptgx2gx'}
\end{eqnarray}

When $a(l,l')$, $b(l,l')$ and $c(l,l')$ are pure numbers as it is
the case for the quasi-linear regime, the integrals in $l$ and
$l'$ factorize and  lead to an expression of the form,
\begin{eqnarray}
\langle \gamma^2(\vx)\vgam(\vx')\rangle&=&
c_{2\,1}\sigma^2_{\kappa}\times\nonumber\\ &&\int \d^2 \vl'
P_{\kappa}(l')W(l'\theta_0)\,e^{\i\vl'.(\vx'-\vx)}\,\vu(\theta_{\vl'})
\label{3ptshear1}
\end{eqnarray}
{where $c_{2\,1}$ is a pure number. It can actually be noted that
$c_{2\,1}$ identifies with the coefficient appearing in the
expression of the reduce joint cumulant,}
\begin{equation}
\langle \kappa^2(\vx)\kappa(\vx')\rangle=
c_{2\,1}\,\sigma^2_{\kappa}\,\langle
\kappa(\vx)\kappa(\vx')\rangle
\end{equation}
in the large separation limit~(Bernardeau 1996). It has actually
been already computed in Bernardeau et al. (1997) in the
quasi-linear regime. It scales like $\Omega_{\rm m}^{-0.8}$ and it
is mainly independent of the spectrum index.

The last integral in (\ref{3ptshear1}) is equal to the
convergence-shear correlation function,
\begin{equation}
\vxi_{\kappa\,\gamma}(\vx')\equiv\int \d^2 \vl'
P_{\kappa}(l')W(l'\theta_0)\,e^{\i\vl'.\vx'}\,\vu(\theta_{\vl'}) \
,
\end{equation}
 and can be computed by noticing that,
\begin{eqnarray}
\int \d^2 \vl'
P_{\kappa}(l')W(l'\theta_0)\,e^{\i\vl'.(\vx'-\vx)}\,\vu(\theta_{\vl'})&=&\nonumber\\
&&\hspace{-5.5cm}-\left(\begin{array}{c}
\partial_{x'}^2-\partial_{y'}^2\\ 2\partial_{x'}\partial_{y'}
\end{array}\right)
\int {\d^2 \vl'\over l'^2}
P_{\kappa}(l')W(l'\theta_0)\,e^{\i\vl'.(\vx'-\vx)}.
\end{eqnarray}
So, if one defines $\xi_{\phi}$ as the potential two-point
correlation function,
\begin{eqnarray}
\xi_{\phi}(x)&=& \int {\d^2 \vl'\over l'^2}
P_{\kappa}(l')W(l'\theta_0)\,e^{\i\vl'.\vx}\nonumber\\ &=&2\pi\int
{\d l'\over l'}P_{\kappa}(l')W(l'\theta_0)\,J_0(l'\,x),
\end{eqnarray}
we get,
\begin{eqnarray}
\int \d^2 \vl'
P_{\kappa}(l')W(l'\theta_0)\,e^{\i\vl'.\vx'}\,\vu(\theta_{\vl'})&=&\nonumber\\
&&\hspace{-5cm}-\left(\begin{array}{c} x'^2-y'^2\\ 2\,x'\,y'
\end{array}\right)
\left[{1\over \vert\vx'\vert^2}\xi''_{\phi}(\vert\vx'\vert)-
{1\over \vert\vx'\vert^3}\xi'_{\phi}(\vert\vx'\vert)\right].
\end{eqnarray}
Therefore, in the case of  a power law spectrum, $P(l)\sim l^n$,
one has for $x\gg \theta_0$
\begin{equation}
\xi_{\phi}(x)\sim x^{-n}
\end{equation}
so that,
\begin{equation}
\left[\xi''_{\phi}(\vert\vx'\vert)- {1\over
\vert\vx'\vert}\xi'_{\phi}(\vert\vx'\vert)\right]
=n(n+2){\xi_{\phi}(\vert\vx'\vert)\over \vert\vx'\vert^2}
\end{equation}
which {is equal to} $-{(n+2)/n}\ \xi_{\kappa}(\vert\vx'\vert)$.
Hence, the 3-point function clearly depends on the slope of the
power spectrum:
\begin{equation}
\langle \gamma^2(\vx)\vgam(\vx')\rangle=-{n+2\over n}\  c_{2\,1}\
\sigma^2_{\kappa}\  \xi_{\kappa}(\vert\vd'\vert)\
\left(\begin{array}{c} \cos(2\theta_{\vd'})\\
\sin(2\theta_{\vd'})
\end{array}\right),
\end{equation}
{where $\vd'$ is the distance vector between $\vx'$ and $\vx$,
$\vd'=\vx'-\vx$, and $\theta_{\vd'}$ is its angle to the first
axis.} It is interesting to notice that, contrary to the
convergence field, the amplitude of the three-point shear function
vanishes (in units of the square of the two-point function) when
$n\to -2$. Although this result is obtained in some specific
limiting configuration it expresses a general trend: when $n$ is
close to $-2$ the shear is dominated by very long
wavelength\footnote{{when $n=-2$ the computation of the variance
of $\kappa$ (or $\gamma$) shows a divergence at $l\to 0$ when it
is computed with the Limber approximation. It means that in the
limit $n\to -2$ the fluctuations of the shear field are dominated
by infinitely long wave modes, much longer than
$\vert\vx-\vx'\vert$. Moreover when $n<-2$ the whole calculation
presented here, which is based on the small angle approximation,
becomes invalid.}}, much longer than $\vx'-\vx$, so that what is
computed here \bold{is the same as} contracted three point
functions that all vanish for symmetry reasons.

For $n\approx -1.5$ and sources at redshift unity, we know from
Bernardeau et al. (1997) that in the quasilinear regime,
\begin{equation}
c_{2\,1} \approx {36.7/ \Omega_{{\rm m}}^{0.8}} \ ,
\end{equation}
{a result that can be obtained from Eq. (\ref{3ptgx2gx'}) with
$a=10/7$, $b=1$, $c=4/7$.} Consequently, observables like
${\langle \gamma^2(\vx)\vgam(\vx')\rangle/\left(
\sigma^2_{\kappa}\, \xi_{\kappa}(\vert\vx'\vert)\right)}$ would
provide alternative ways for measuring the cosmic density
parameter $\Omega_{{\rm m}}$. They do not require mass
reconstruction but  still require some filtering which, for the
reasons mentioned in the beginning we would like to avoid.

\subsubsection{Computation of $\left\langle
\left(\vec\gamma(\vx_1).\vec\gamma(\vx_2)\right)\vgam(\vx')\right\rangle$}

In the previous paragraph the calculations were tractable without
strong hypothesis on the shape of the bispectrum. Here, we explore
more generic geometrical cases, so more specific assumptions on
the bispectrum are necessary to carry out analytical computations.
We assume it follows the prescription usually adopted in the
strongly non-linear regime, that is the coefficient $Q_{\kappa}$
introduced in Eq. (\ref{Qkappa}) is constant (but depends on the
cosmological model). We do expect this is a valid approximation at
the scales we are interested in (below 5') and in any case
 the results presented in the following do not critically  rely
on this assumption as it is the case for the skewness of the
convergence.

A quantity like $\left\langle
\left(\vgam(\vx_1).\vgam(\vx_2)\right)\,\vgam(\vx')\right\rangle$
is expected to behave as $\langle\vgam^2(\vx)\,\vgam(\vx')\rangle$
when $\vert \vx_1-\vx_2\vert\ll\vert\vx'-(\vx_1+\vx_2)/2\vert$.
{In this limit its expression is given by,
\begin{eqnarray}
\left\langle \left(\vgam(\vx_1).\vgam(\vx_2)\right)\,
\vgam(\vx')\right\rangle&=&\int\d^2\vl\,\d^2\vl'
\,P_{\kappa}(l)\,P_{\kappa}(l')\times
\nonumber\\
&&\hspace{-2cm}\,Q_{\kappa}(\vl,\vl')\,
e^{\i\vl.(\vx_2-\vx_1)+\i\vl'.(\vx'-\vx_1)}\times\nonumber\\
&&\hspace{-2cm}
\cos(2\theta_{\vl+\vl'}-2\theta_{\vl'})\,\vu(\theta_{\vl'})+\left\{\vx_1
\leftrightarrow \vx_2\right\} \label{gam1gam2gamp}
\end{eqnarray}
Here the filtering effects can be ignored since all points are
taken at finite distance.}

{In case of a simple bispectrum -- with only a non-zero monopole
term and $b=c=0$ -- and with the help of the expansion
(\ref{cos2tmtp}) the integrals over the angle $\theta_{\vl}$ and
$\theta_{\vl'}$ in (\ref{gam1gam2gamp}) can be computed
explicitly. It leads to an expression of the form, }
\begin{eqnarray} \langle \left(\vgam(\vx_1).\vgam(\vx_2)\right)\,
\vgam(\vx')\rangle=
c_{2\,1}\times\nonumber\\
&&\hspace{-4cm}\xi_{\kappa}(\vert\vx_2-\vx_1\vert)\
\xi_{\kappa\,\gamma}(\vert\vd'\vert)\ \left(\begin{array}{c}
\cos(2\theta_{\vd'})\\ \sin(2\theta_{\vd'})
\end{array}\right),
\label{s3farpair}
\end{eqnarray}
{where $\vd'=\vx'-(\vx_1+\vx_2)/2$. It generalizes the result of
Eq. (\ref{3ptshear1}).}

This result can be simply interpreted: an excess of shear at a
given position is most likely associated with a mass overdensity,
so that the shear at finite distance from the pair points is
preferentially tangential. This non-zero three point function is
therefore simply associated with the usual skewness. It directly
comes from the relative excess of overdensity regions compared to
low density areas.

The situation becomes more complex when the assumption $\vert
\vx_1-\vx_2\vert\ll\vert\vx'-(\vx_1+\vx_2)/2\vert$ is dropped.
Another restricting case can however be investigated, when $\vert
\vx'-\vx_1\vert\ll\vert\vx_1-\vx_2\vert$. {In this case the two
dominant contributions are}
\begin{eqnarray}
\left\langle \left(\vgam(\vx_1).\vgam(\vx_2)\right)\,
\vgam(\vx')\right\rangle&=&\int\d^2\vl\,\d^2\vl'
\,P_{\kappa}(l)\,P_{\kappa}(l')\,Q_{\kappa}(\vl,\vl')\times
\nonumber\\
&&\hspace{-3cm}\,\,\left[
e^{\i\vl.(\vx_2-\vx_1)+\i\vl'.(\vx'-\vx_1)}
\cos(2\theta_{\vl+\vl'}-2\theta_{\vl})\,\vu(\theta_{\vl'})+
\right.\nonumber\\
&&\hspace{-2.5cm} \left.e^{\i\vl.(\vx'-\vx_1)+\i\vl'.(\vx'-\vx_2)}
\cos(2\theta_{\vl}-2\theta_{\vl'})\,\vu(\theta_{\vl'+\vl})\right]
\label{gam1gampgam2}
\end{eqnarray}
Then similar calculations can be performed, again assuming the
bispectrum simply factorizes in terms of the power spectrum as in
the nonlinear regime. It leads to the expressions,
\begin{eqnarray}
\left\langle \left(\vgam(\vx_1).\vgam(\vx_2)\right)\,
\vgam(\vx')\right\rangle&=&
c_{2\,1}\ \xi_{\kappa\,\gamma}\left(\vert\vx_2-\vx_1\vert\right)\times\nonumber\\
&&\hspace{-3.5cm}\left[ \xi_{\kappa}(\vert\vx'-\vx_1\vert)
\left(\begin{array}{c}1\\ 0 \end{array}\right)\ +\right.\nonumber\\
&&\hspace{-3.5cm}\left.  \left(\xi_{\rm tt}\left(\vert\vx'-\vx_1\vert\right)
-\xi_{\rm rr}\left(\vert\vx'-\vx_1\vert\right)\right)
\left(\begin{array}{c}\cos(4\theta_{\vx'-\vx_1})\\
\sin(4\theta_{\vx'-\vx_1})\end{array}\right) \right]
\label{s3nearpair}
\end{eqnarray}
if the $\vx_2-\vx_1$ vector is along the first direction. In this
expression we have decomposed the shear two-point correlation
function into the tangential part $\xi_{\rm tt}$ {(i.e. the
correlation function of the shear components along the
$\vx_2-\vx_1$ direction) and the radial part $\xi_{\rm rr}$}. The
relative importance of the 2 terms depend on the power spectrum
shape, e.g.
\begin{equation}
Q_2\equiv{\xi_{\rm tt}-\xi_{\rm rr}\over
\xi_{\kappa}}={(n+2)(n+4)\over (n-2)n}.
\end{equation}
This shape is actually a direct transcription of the  behavior of
the shear two-point functions \bold{(for a component $\gamma_1$ of
$\vgam(\vx_1)$ along a fixed direction)},
\begin{eqnarray}
\langle \gamma_1(\vx_1)\ \vgam(\vx')\rangle&=& \left[
\xi_{\kappa}(\vert\vx'-\vx_1\vert) \left(\begin{array}{c}1\\ 0
\end{array}\right)\ +\right.\nonumber\\ &&\hspace{-2.5cm}\left.
\left(\xi_{\rm tt}(\vert\vx'-\vx_1\vert) -\xi_{\rm
rr}(\vert\vx'-\vx_1\vert)\right)
\left(\begin{array}{c}\cos(4\theta_{\vx'-\vx_1})\\
\sin(4\theta_{\vx'-\vx_1})\end{array}\right) \right].
\end{eqnarray}
Nonetheless these patterns are indicative of the physical effects
one might want to look for. The {pseudo-vector field} structure of
the three-point function {displays a pattern which} corresponds to
the superposition of a uniform field and a specific mass-dipole
contribution.  The relative weight of the two terms depends on the
power law index.  When $n$ is close to -2, structures are mostly
dominated by long-wavelength modes and the shear patterns are
aligned (in other words $\xi_{\rm tt}$ and $\xi_{\rm rr}$ are
equal). In contrasts when $n$ is close to 0, structures are given
by point-mass Poisson distribution (and $\xi_{\rm tt}$ and
$\xi_{\rm rr}$ are opposite to each-other).

Plots on Fig. \ref{shears3nm1p5} show the shear patterns near pair
points for different power law indices. One can see that for $n
\le -1$ the shear pattern around the points is mostly uniform.

\subsection{Semi-analytical results}

\begin{figure}
\begin{tabular}{c}
{\centering\resizebox*{7.5cm}{7.5cm}{\includegraphics{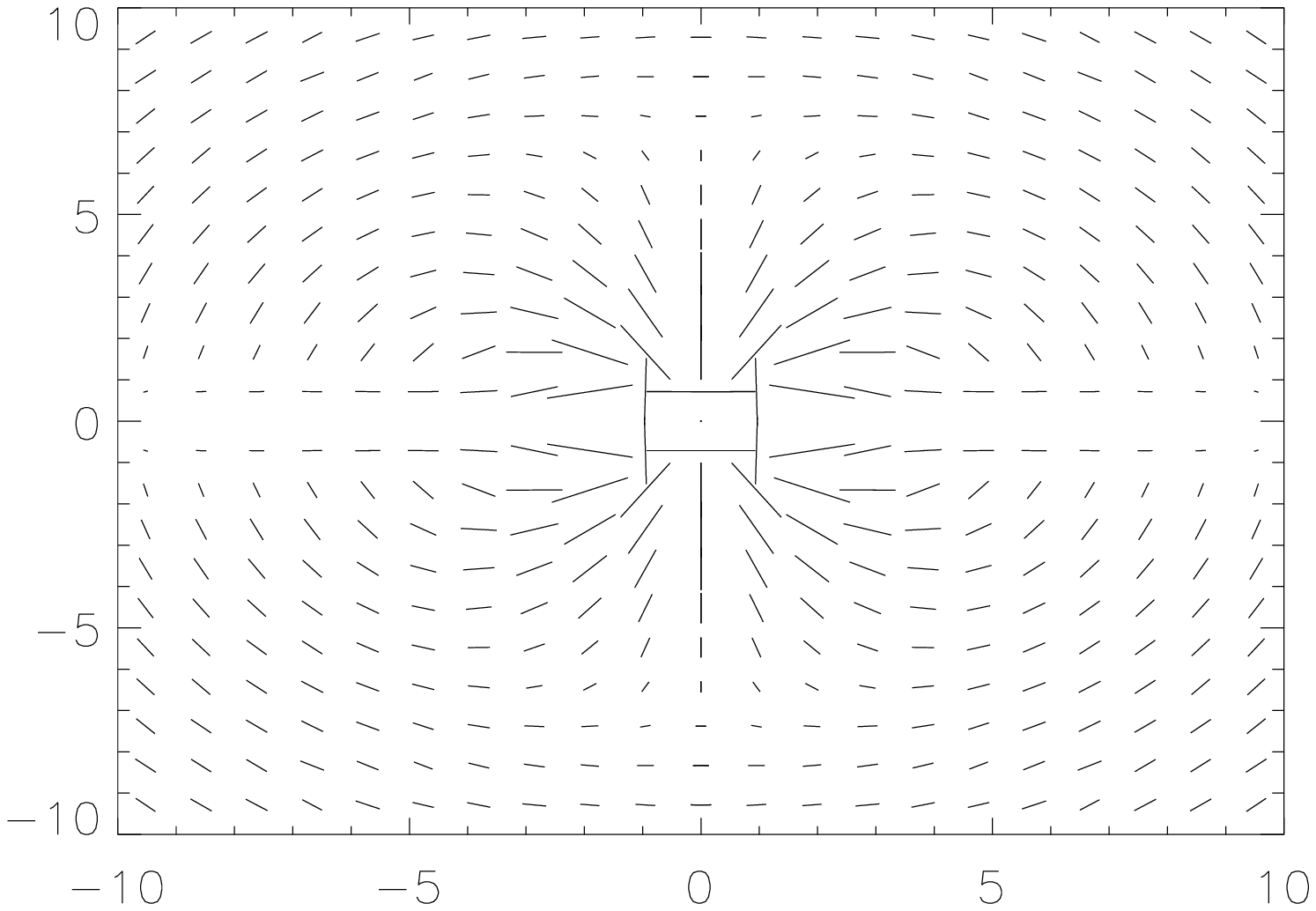}}}\\
{\centering\resizebox*{7.5cm}{7.5cm}{\includegraphics{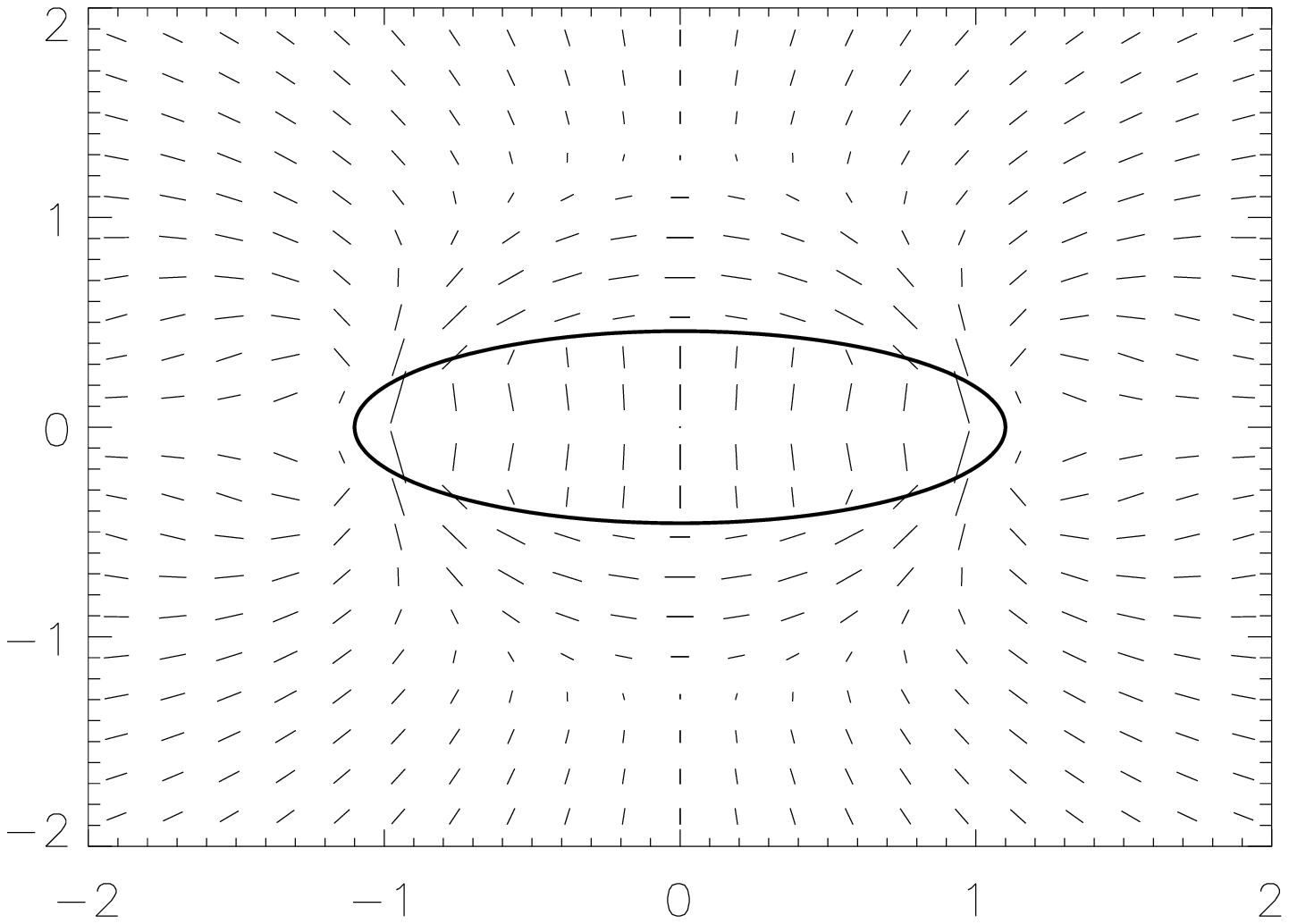}}}
\end{tabular}
\caption{The shear pattern obtained by a numerical integration in
case of a power spectrum index $n=-1$ and for a three-point
function that obeys the hierarchical ansatz, $Q_{\kappa}=$
constant in Eq. (\ref{Qkappa}). { The pair points are at position
$(-1,0)$ and $(1,0)$ in both panels. The only difference in the
two panels is the scale at which the pattern is drawn.}}
\label{patternm1}
\end{figure}

The exact analytical results presented in the previous paragraphs
only correspond to  simple cases. They provide insightful
descriptions of the behavior of the shear three-points function
but have a limited practical interest if their exact validity
domain is unknown. More general results can  be obtained, but only
for specific cosmological models through numerical computation and
ray-tracing simulations.

\begin{figure}
{\centering\resizebox*{8cm}{!}{\includegraphics{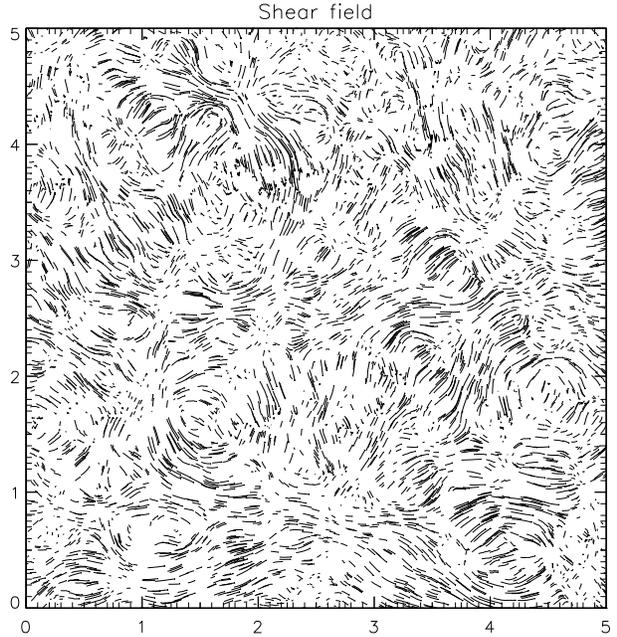}}}
\caption{Example of shear map obtained from a numerical simulation
(open CDM model, for an 5x5 square degree survey, van Waerbeke et
al. 1999)  } \label{shear_map}
\end{figure}

Fig. \ref{patternm1} shows the shear pattern expected for a power
law spectrum of index $n=-1$ and for a matter three-point function
shape given by the non-linear regime.
%In general the computations of
%these quantities are complicated because of the very slow convergence
%of the integrals, specially for power law models. For $n=-1$, the
%computations are made easier because the integral over the
%wave-vector length can be done explicitly.
The patterns observed around the pair points (located at positions
[-1,0] and [1,0]) are rather complicated. In particular the
circular shape expected around the pair points (according to the
results of sect. 2.2.2) are observed only at very large distances
(more that 5 times the pair separation). At separation comparable
to the pair distance the patterns \bold{increase complexity}, with
a substantial part of the area with no significant correlation.
The most striking feature is the quasi-uniform shear orientation
along the segment joining the two-points. This effect which is
indeed expressed in Eq. (\ref{s3nearpair}), where a uniform
component is explicitly predicted,  clearly strengthens in between
the pair points. Such a feature might appear somewhat surprising
but {a close inspection of synthetic shear maps (Fig.
\ref{shear_map}) indeed reveals} \bold{a lot of highly contrasted
clumpy regions surrounded by strong coherent shear patterns
primarily oriented transversely to directions between lumps.}

The central pattern is the strongest and the most typical feature
of the three-point correlation map and should be the easiest
detectable one in the data.  The previous analytical results
suggest that such a structure is expected to hold for power law
indices between $-2$ and $-1$ and should then be robust enough to
be used as a detection tool of non-gaussianities. More detailed
analysis performed along this idea are presented in the following.

\section{Comparison with numerical simulations}

\begin{figure}
\begin{tabular}{cc}
{\centering\resizebox*{4cm}{4cm}{\includegraphics{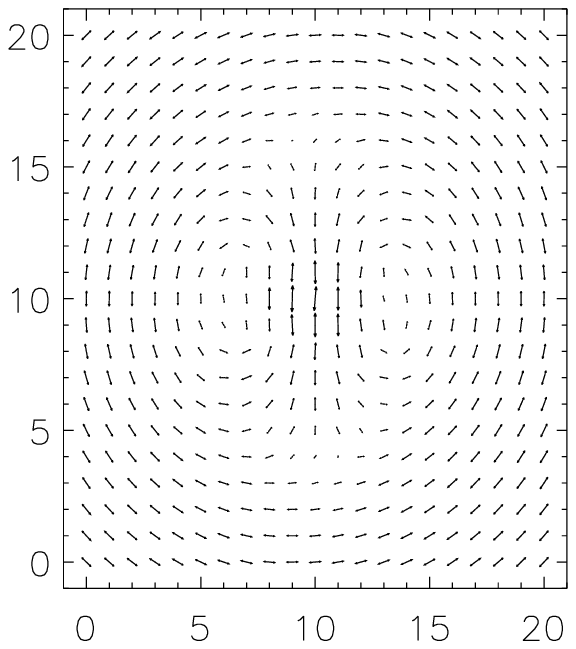}}}&
{\centering\resizebox*{4cm}{4cm}{\includegraphics{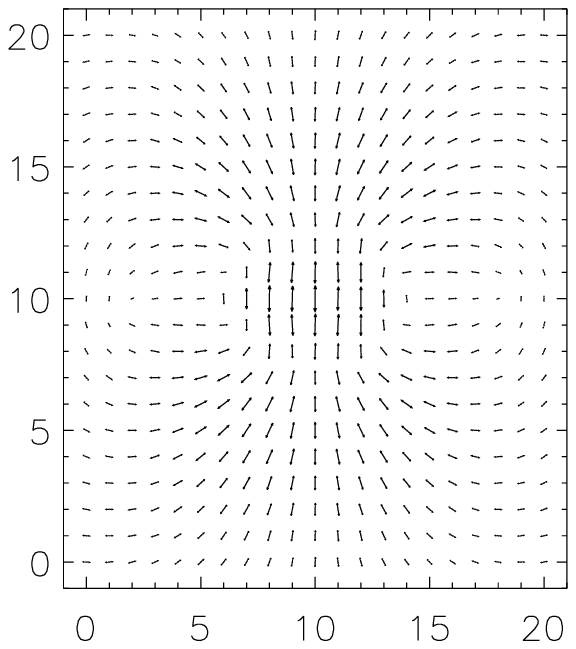}}}\\
{\centering\resizebox*{4cm}{4cm}{\includegraphics{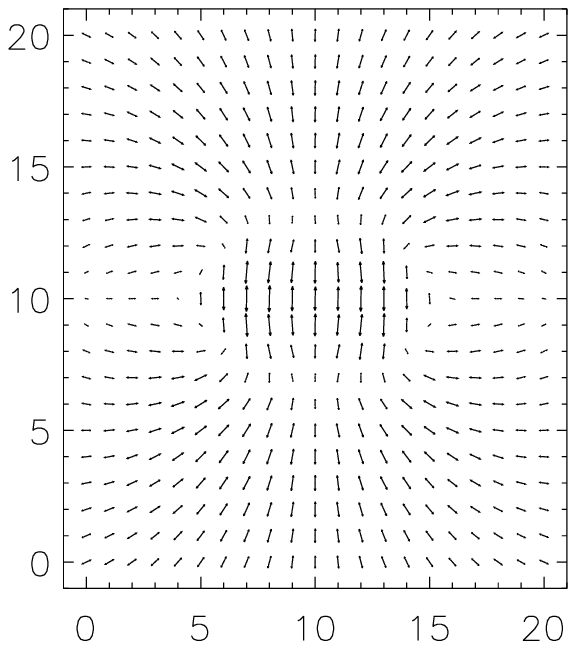}}}&
{\centering\resizebox*{4cm}{4cm}{\includegraphics{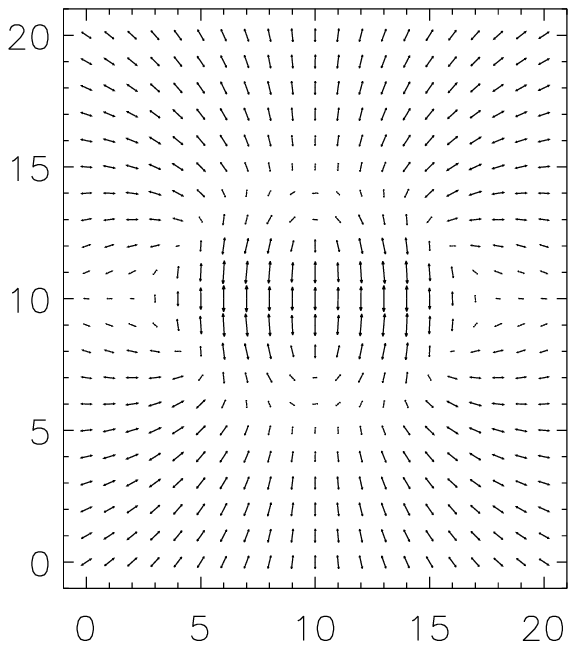}}}\\
\end{tabular}
\caption{The shear pattern of $\left\langle
(\vec\gamma(\vx_1).\vec\gamma(\vx_2))\vgam(\vx')\right\rangle$
measured in ray-tracing simulations (Jain et al., 2000) for
increasing pair separation $\vert\vx_2-\vx_1\vert$. The
separations are, from top to bottom and left to right, 2, 4, 6,
and 8 in plot units (1 unit corresponds to about 10''). The pair
points are along the horizontal axis.} \label{pattern_simul}
\end{figure}

The numerical simulations we use are described in Jain et al.
(2000). The cosmological model is an open Universe ($\Omega_{{\rm
m}}=0.3$,
$\Omega_{\Lambda}=0$) with a Cold Dark Matter power spectrum
($\Gamma=0.21$) and a normalization $\sigma_8=0.85$. The sources
are located at redshift unity and the simulation area covers about
11 square degrees with a resolution of $0.1$ arcmin.

The shear patterns for pair points at increasing separation is
shown on (Figure \ref{pattern_simul}). A visual inspection of
their morphology and strength confirms that the uniform shear
pattern within an ellipse that encompasses the pair points (as
described in the next section) is likely an optimum way to extract
non-gaussian signal. When the separation is small, the overall
circular shear pattern is clearly visible. When the separation
increases the shear appears uniform in the neighborhood of the
segment joining the pair points, and is mostly radial at finite
distance. We have already seen that these results might somewhat
be dependent on the power spectrum index. For this simulation the
index varies from $-1.3$ to $-1$ and it is thus natural that the
patterns look like those obtained in case of a power-law model
$n=-1$.

\section{Improved measurement strategies}

In this section we compare the measurements made in mock catalogs
that mimic a large number of observational effects with
 different input models. We use these results to
develop different survey strategies adapted to real data set.

\subsection{Mock catalogues}

The mock catalogues are generated from simulated sky images
following the procedure described in Erben et al. (2001). The only
difference here is that the galaxies are lensed according to a
realistic cosmic shear signal, using ray-tracing simulations (Jain
et al. 2000) instead of having a constant shear amplitude as in
Erben et al. 2001. The galaxies are analyzed exactly in the same
way as real data, following the procedure described in van
Waerbeke et al. (2000, 2001b). In particular, the mock catalogues
contain the main three features encountered in the actual surveys:
\begin{itemize}
\item galaxy intrinsic shape fluctuations; \item masks; \item
noise from galaxy shape measurements and systematics from PSF
corrections \ ,
\end{itemize}
\begin{figure}
\centering\resizebox*{6.cm}{5cm}{\includegraphics{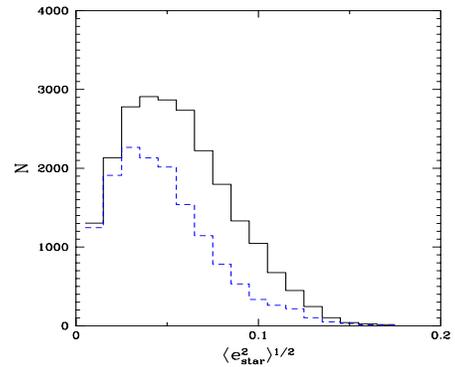}}
%\centering{\includegraphics{star_anisotropy.ps}}
\caption{Histogram of the Point Spread Function anisotropy of the
stars in the simulated images (solid line) and in the
VIRMOS-DESCART survey (dashed line).} \label{psf_aniso}
\end{figure}
and the simulated images reproduce observational conditions
matching our data set (PSF anisotropy, limiting magnitude,
luminosity functions, galaxy and star number densities, intrinsic
ellipticity...). The PSF anisotropy is even larger than
the one in our data set, as shown in Figure \ref{psf_aniso}, but
it is uniform \footnote{However, we do NOT assume a uniform PSF
for the PSF correction, as for the real data, we fit the PSF with a
2D second order polynomial}. We
used two ray-tracing simulations from Jain et al. (2000): one is
OCDM, as described in Section 3, and the other is a $\tau$CDM with
$\Gamma=0.21$ and $\Omega_{{\rm m}}=1$.
 For each simulation we produced $11$ square degrees of
simulated sky images containing roughly $30$ galaxies per
arcmin$^2$, with a pixel size of $0.2$ arcsec. The galaxies are
lensed following the ray-tracing shear before generating the image
\footnote{That is the image of the galaxies are not lensed. We
first lens  the catalogue of source galaxies and  then use it to
generate the sky images}. In addition to the galaxy mock
catalogues, we produced a reference catalogue containing only the
cosmic shear values. We can therefore study separately the effect
of masks, the ellipticity Poisson noise, real noise and
systematics, when compared to the reference catalogue results.

\subsection{Detection strategy}

\begin{figure*}
\begin{tabular}{cc}
{\centering\resizebox*{8.5cm}{6cm}{\includegraphics{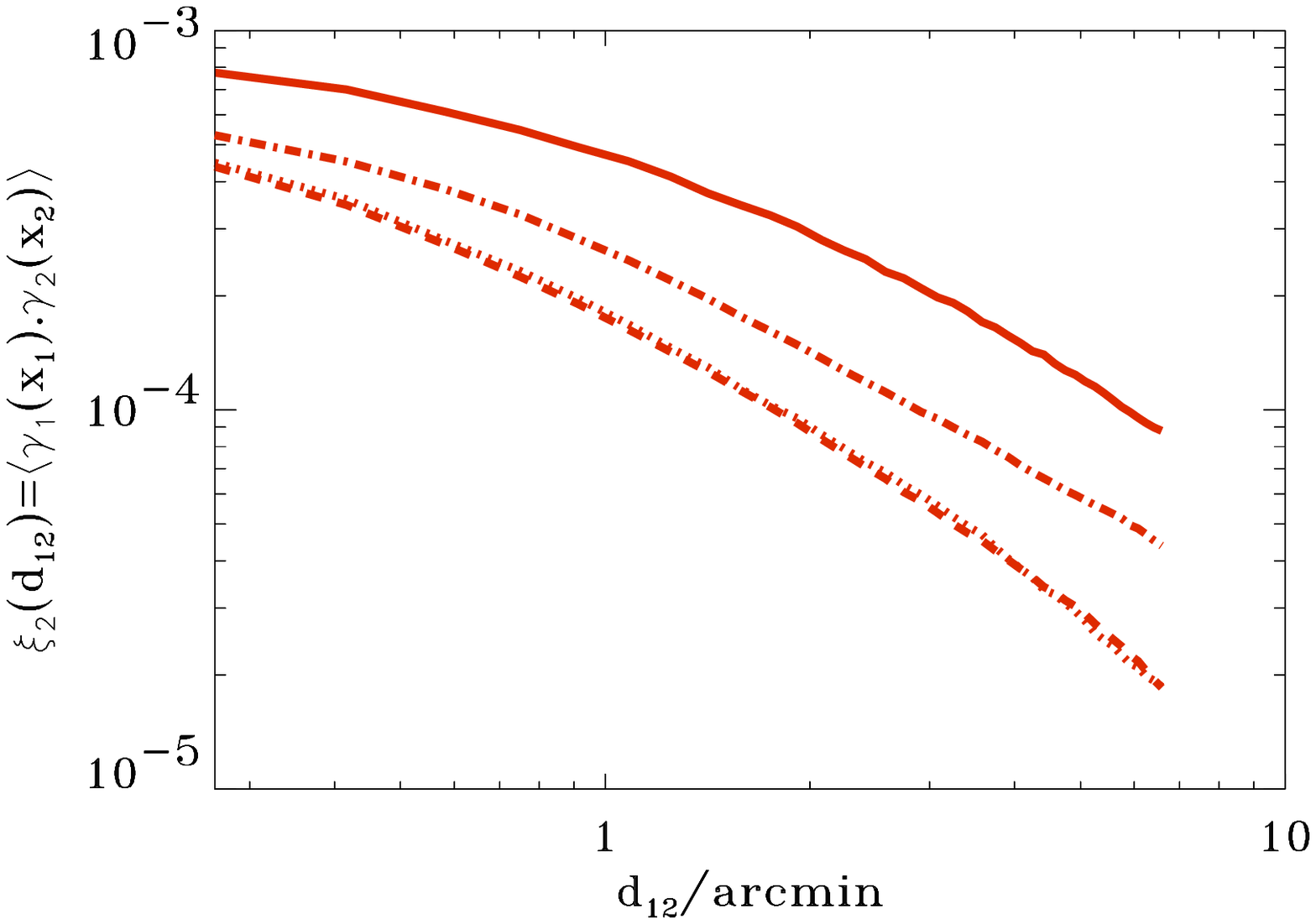}}}&
{\centering\resizebox*{8.5cm}{6cm}{\includegraphics{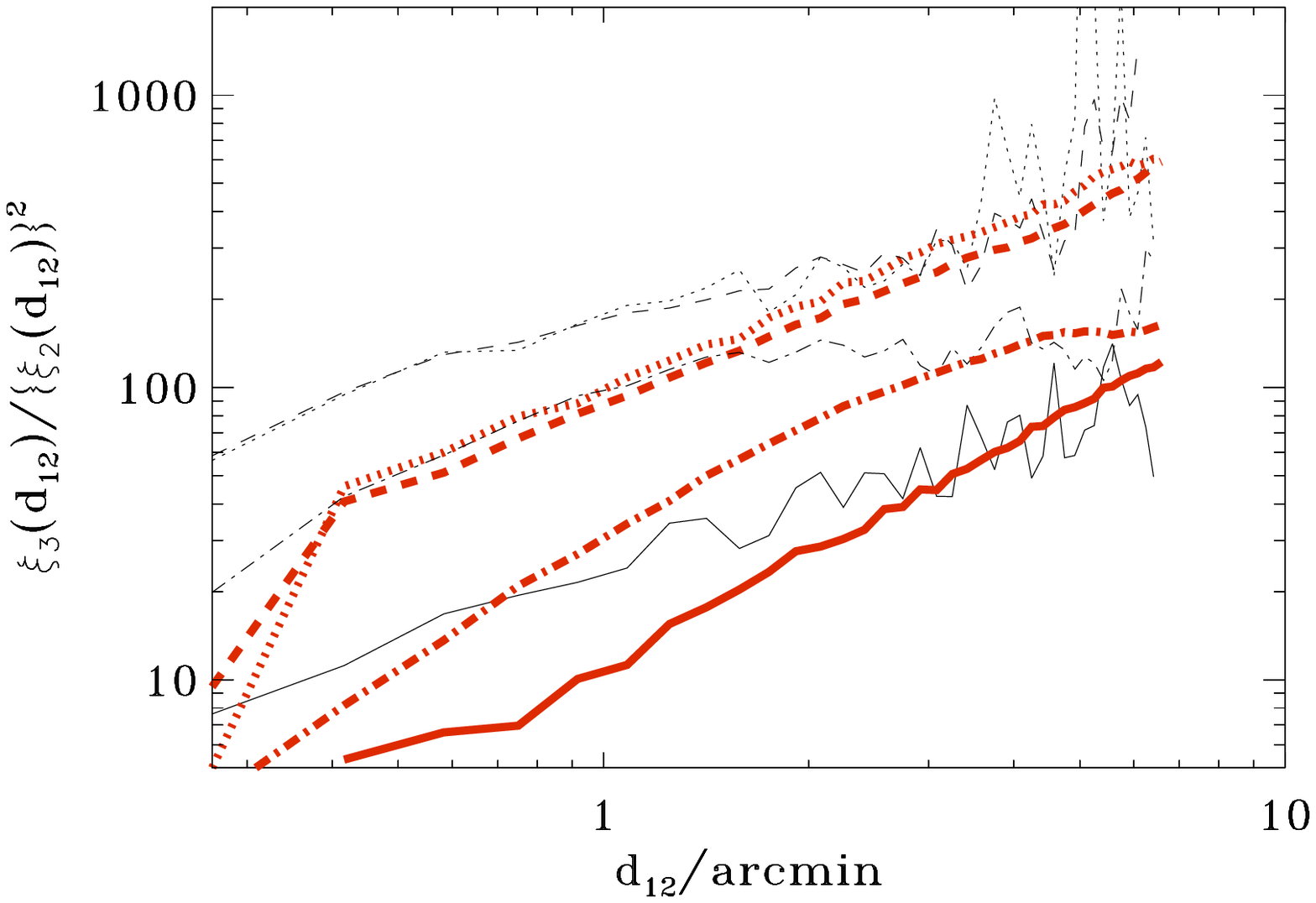}}}\\
\end{tabular}
\caption{Results for the measured value of the two-point function
and the reduced three-point function (as defined in Eq.
\ref{xi3_ellip}) as a function of the pair distance for an Open
CDM model (thick dashed lines) a $\Lambda$CDM (thick dot-dashed
lines) and a $\tau$CDM model (thick solid lines). Results obtained
in an Open CDM models with masks is shown as thick dotted lines.
Galaxies in the catalogs have no intrinsic ellipticities. The
measurements are made in bins of width 10''. The thin lines of the
right panel corresponds to $200\,(n+2)/(-n)$, $150\,(n+2)/(-n)$
and $60\,(n+2)/(-n)$ for the open, $\Lambda$ and $\tau$ CDM case
respectively, where $n$ is the power spectrum index measured from
the slope of the corresponding two-point correlation functions. }
\label{xi3exact}
\end{figure*}

The adopted strategy is to measure {the two component quantity}
\begin{eqnarray}
\overline{\xi_3}(\vert\vx_1-\vx_2\vert)= \int_{\rm
ellipse}{\d^2\vx'\over V_{\rm ellipse}} \langle
\left(\vgam(\vx_1).\vgam(\vx_2)\right)\, \vgam(\vx')\rangle
\label{xi3_ellip}
\end{eqnarray}
for different separations. {The {\it bar} means we perform the
average of $\langle\left(\vgam(\vx_1).\vgam(\vx_2)\right)\,
\vgam(\vx')\rangle$ within the area defined by $V_{\rm ellipse}$.
} The ellipse properties are  defined to cover an area that
encompasses the pair points in which the orientation of the shear
pattern is expected to be uniform. To be more specific, after some
trials we have obtained a good balance between the
 signal intensity and the shot noise amplitude when the
ellipse has a $0.4$ eccentricity, the pair points being its foci
(see on Fig. \ref{patternm1}).

The computation time can be reduced by building first a Delaunay
triangulation (Delaunay, 1934; \bold{see also van de Weygaert,
1991}) of the survey so that the neighbor list of any galaxy in
the survey can be easily (and quickly) constructed. {The quantity
$\overline{\xi_3}(\vert\vx_1-\vx_2\vert)$ is a pseudo-vector, but
its second component, e.g. the one corresponding to elongation at
45 degrees to $\vx_1-\vx_2$, vanishes in average for symmetry
reasons. Therefore we only have to compute one component, which we
define as,}
\begin{equation}
\overline{\xi}^{\rm {\ t}}_3(d_{ij}) ={\sum_{ijk}
w_i\,w_j\,w_k\,(\ve_i.\ve_j)\ e^{(ij)}_k \over \sum_{ijk}
w_i\,w_j\,w_k }
\end{equation}
where $\ve_i$ are the galaxy ellipticities, $e^{(ij)}_k$ is the
tangential (e.g. first) component, with respect to the segment
$\vx_j-\vx_i$, of the  ellipticity of the galaxy of index $k$,
$w_i$ are weights associated with each galaxy. The sums are made
under the following constraints:
\begin{itemize}
\item $d_{ik}>d_{min}$, $d_{jk}>d_{min}$ so that
close pairs are excluded to avoid spurious small angular scale
signals (van Waerbeke et al. 2000) and
\item  $d_{ik}+d_{jk}<1.1\,d_{ij}$ so
that galaxy $k$ is within the afore defined ellipse and the sum is
splitted in bins according to the $d_{ij}$ distance.
\end{itemize}

We then plot the reduced three point function, that is
$\overline{\xi}^{\rm {\ t}}_3$ in units of $[\xi_2(d_{ij})]^2$,
\begin{equation}
\xi_2(d_{ij})={\sum_{ij} w_i\,w_j\,(\ve_i.\ve_j) \over \sum_{ijk}
w_i\,w_j}.
\end{equation}

The noise is estimated by computing the r.m.s. of the estimator in
small individual bins (for which the noise dominates the signal),
and then rebinned in the final, larger, bins (as described in Pen
et al., 2002). The noise of the reduced three-point function is
computed assuming that the noise of the second and the third
moments are uncorrelated. This might not be exact but {the
consequence of this assumption is insignificant since the main
contribution to the noise of the ratio is in general the
three-point moment}.

\subsection{Behavior of the reduced three-point function}

In  Fig. \ref{xi3exact} we show the behavior of the reduced
three-point function in the simulations. These simulations do not contain
  any noise,
except that the shear field is measured in discrete randomly
placed points.  It is found to be a slowly growing function of
scale. This might first appear surprising (this ratio is expected
to be constant for a constant $Q$) but this result should be
examined in the light of  equation (\ref{s3farpair}) and the
subsequent comments: at small scale $n$ decreases, therefore it is
expected that the reduced three-point function also decreases.
Indeed, as demonstrated on the right panel of Fig. \ref{xi3exact},
the reduced ratio behaves approximately like $(n+2)/n$ {at least
for scale above 2' with a multiplicative coefficient ($200$, $150$
and $60$ for respectively an open, $\Lambda$ and $\tau$ CDM model)
that roughly correspond to the values of the convergence skewness
$s_3$ at small scale in those models~(Hui 1999).}

On this plot the effects of masks on these quantities are shown as
dotted lines. As expected its effect on the two-point function is
null. Its effect on the three-point function as measured here is
not totally absent (different configurations taken into account in
Eq.(\ref{xi3_ellip}) may have different weights when masks are
taking into account). It appears however that it has a  negligible
effect compared to the other sources of noise.

Fig. \ref{xi3noise} also shows the amplitude of the cosmic
variance to be expected in an 11 square degree survey for the
reduced three-point function and in case of an Open CDM model
(dotted dashed lines). These error bars have been obtained from
the results obtained in 7 realizations of the same model. They
show that 30\% fluctuation is to be expected in the signal. It is
to be noted also that the error bars \emph{are} correlated in the
different bins.

\subsection{Optimization and noise effects}

\begin{figure}
{\centering\resizebox*{8.5cm}{6cm}{\includegraphics{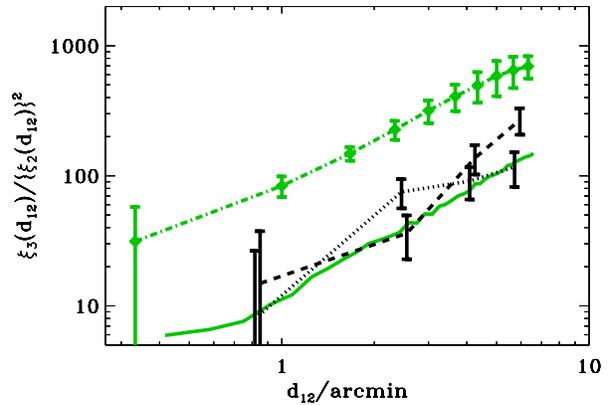}}}
\caption{ The reduced three-point function as in previous figure
for different noise types in the catalogs. Error-bars for the open
CDM model (dotted-dashed line) correspond to a \emph{cosmic
variance} estimation. The plot also compares the case of exact
ellipticities for a $\tau$CDM model (solid lines) with the case
where galaxies have intrinsic ellipticities (dotted lines, the
measurement is made in 4 independent bins) and the case where
noises in the shape determination are included (after PSF
corrections, etc..., dashed lines, 4 bins). Error bars correspond
here to \emph{measurement errors}, not to the cosmic variance, and
are therefore uncorrelated.} \label{xi3noise}
\end{figure}

A proper choice of weights $w_i$ is essential for getting a good
signal to noise ratio. We have found that the introduction of a
cut-off in the galaxy ellipticity distribution,
\begin{equation}
w_i \sim \exp\left(-{e_i^2\over 2\,e_0^2}\right) \ ,
\label{cutoff}
\end{equation}
improves the signal detection but may also affect its amplitude.
In figure \ref{xi3noise} a cut-off is introduced with $e_0=0.5$ in
order to improve the S/N ratio.  If the measured shear is simply
the sum of the intrinsic shear and some noise,
\begin{equation}
\ve_i=\vgam(\vx_i)+\epsilon_i
\end{equation}
then the cut in $e$ translates in a cut in $\gam$ that can be
simply described. The result depends on the shape of
the intrinsic ellipticity PDF, $P(\eps)$,
\begin{equation}
\vgam_i^{\rm cut}={\int \d^2\veps\ (\vgam(\vx_i)+\epsilon_i)\,
P(\eps)\,\exp\left(-{(\vgam_i+\veps)^2\over 2\,e_0^2}\right)\over
\int \d^2\veps \, P(\eps)\,\exp\left(-{(\vgam_i+\veps)^2\over
2\,e_0^2}\right)}.
\end{equation}
If the cut-off value $e_0$ is large enough compared to the typical
excursion of cosmic shear values, then this equation can be
linearized in $\gam$ and one finds that,
\begin{equation}
\vgam_i^{\rm cut}=f_c\,\vgam(\vx_i)+\ldots \label{fcdef}
\end{equation}
with
\begin{equation}
f_c={\int \eps\,\d\eps \left(1-{\eps^2\over 2\,e_0^2}\right)\,
P(\eps)\,\exp\left(-{\eps^2\over 2\,e_0^2}\right)\over \int
\eps\,\d\eps \,P(\eps)\,\exp\left(-{\eps^2\over 2\,e_0^2}\right)}.
\end{equation}

For the adopted ellipticity distribution and for $e_0=0.5$, the
value of $f_c$ is about 0.8. Moreover the cut-off is large enough
to have  no impact on the non-gaussian properties of the shear
field (e.g. sub leading order terms in Eq.(\ref{fcdef}) have a
negligible effect).

These numerical investigations {are useful for exploring} the
effects of various noises. As expected the intrinsic ellipticities
increase the noise, but do not bias the result. For $11$
square-degrees surveys the signal to noise for the detection in a
$\tau$CDM model is still comfortable (see Fig. \ref{xi3noise}).
Masking effects are also rather mild. They bias very weakly the
results (see Fig. \ref{xi3exact}) and only slightly deteriorate
the signal to noise ratio when intrinsic ellipticities are taken
into account (because the number of available pairs or triplets is
smaller).

Moreover, when similar measurements are made in mock catalogues
that incorporate realistic noises we described above, we recover
the expected signal with  still a reasonable signal to noise
ratio: it is still larger than 3 in 3 independent bins (see Fig.
\ref{xi3noise}). {Note} that the S/N ratio for the reduced
skewness is not only sensitive to the amplitude of the three point
function, but also to the amplitude of the 2-point function.
\bold{On the lower curves of Fig. \ref{xi3noise}, {only} the noise
coming from the three-point function measurement is shown. Its
amplitude}is roughly independent on the cosmological model (it
scales basically like the inverse square root of the number of
triplets available).

\section{Conclusion}

We have presented various geometrical patterns that the shear
three-point correlation function is expected to exhibit. Its
dependence with the cosmological parameters is found to be similar
to that of the skewness of the convergence field, opening an
alternative way to break the degeneracy between the amplitude of
the density fluctuations, $\sigma_8$, and the density parameter of
the Universe $\Omega_{{\rm m}}$. However, the reduced three-point
function of the shear is more dependent on the power spectrum
index than the skewness. In particular it is expected to vanish
when the index gets close to $-2$. Numerical investigations have
nonetheless proved the shear patterns to be robust enough to
provide a solid ground for the detection of non-gaussian
properties in cosmic shear fields.

We proposed a detection strategy that has been tested in mock
catalogues that include realistic noise structures such as
residual systematics and PSF anisotropy as seen in the real data.
 The quality of the PSF correction is always good enough to
provide an accurate measurement of the shear three-point function.
Since the mock catalogues were designed to reproduce the
characteristics of the current VIRMOS-DESCART lensing survey, we
conclude that non-gaussian signal should be detectable in this
data set. The result of our investigations is presented in another
paper (Bernardeau et al. 2002).

Beyond the detection, the scientific exploitation of the 3-point
function for cosmology also depends on our ability to overcome
other important difficulties. For instance, we found the cosmic
variance amplitude to be of the order of 30\% of the signal on the
reduced three-point function, in agreement with previous studies
made for the convergence. Other issues regarding source
clustering, source redshift uncertainties, and intrinsic alignment
of galaxies have not been considered here.

The measurement of non-gaussian signatures in lensing surveys is
of course of great interest because it provides an independent
measure of the mean mass density of the Universe in addition to
test the gravitational instability paradigm which lead to large
scale structures. It is likely that the analysis of cosmological
non-gaussian signatures will be one of the major and most
promising goals of emerging dedicated lensing
surveys\footnote{Such as the Canada-France-Hawaii Telescope Legacy
Survey , http://www.cfht.hawaii.edu/Science/CFHLS/} that take
advantage of panoramic CCD cameras of the MEGACAM generation
(Boulade et al 2000).

{ \acknowledgements  We thank B. Jain for providing his
ray-tracing simulations, and the referee P. Schneider for a very
usefull and detailed report. This work was supported by
the TMR Network ``Gravitational Lensing: New Constraints on
Cosmology and the Distribution of Dark Matter'' of the EC under
contract No. ERBFMRX-CT97-0172. The numerical calculations were
partly carried out on MAGIQUE (SGI-O2K) at IAP. The simulated
sky images used in this work are available upon request (60Gb).
FB thanks IAP for hospitality.}

\end{document}